\magnification=1200
\hsize=17.8truecm
\vsize=23.8truecm
\hoffset=-0.35truein

\hyphenation{non-or-thog-o-nal}

\dimen\footins=7truecm

\font\footrm=cmr10 at 10truept
\font\footsl=cmsl10 at 10truept
\font\footbf=cmbx10 at 10truept
\font\footit=cmti10 at 10truept

\newdimen\rulewidth
\rulewidth=0.8truept

\newdimen\halfrulewidth
\halfrulewidth=\rulewidth
\divide\halfrulewidth by 2

\newdimen\extra
\extra=1.2cm

\newdimen\halfextra
\halfextra=\extra
\divide\halfextra by 2

\newdimen\quarterextra
\quarterextra=\extra
\divide\quarterextra by 4

\newdimen\boxhsize
\boxhsize=\hsize
\divide\boxhsize by 2
\advance\boxhsize by -\halfrulewidth
\advance\boxhsize by -\halfextra

\newskip\regbaselineskip
\regbaselineskip=14.5pt plus 0.5pt minus 0.5pt
\baselineskip=\regbaselineskip

\newskip\boxbaselineskip
\boxbaselineskip=14pt 

\newskip\figurebaselineskip
\figurebaselineskip=12.5pt plus 0.5pt minus 0.5pt

\newskip\footbaselineskip
\footbaselineskip=12.5truept plus 0.5truept minus 0.5truept

\nopagenumbers
\raggedbottom
\parindent=1.0cm
\parskip=0.3cm plus 1pt minus 0.5pt

\abovedisplayskip=12pt plus 3pt minus 5pt 
\belowdisplayskip=12pt plus 3pt minus 3pt 
\abovedisplayshortskip=4pt plus 2pt
\belowdisplayshortskip=7pt plus 2pt minus 1pt

\newcount\secno
\secno=0
\newcount\subsecno
\subsecno=0

\newdimen\secindent
\secindent=0.55cm

\def\section#1\par{\bigbreak
                   \subsecno=0
                   \global\advance\secno by 1
                   \noindent
                   \hbox to \secindent{\the\secno.\hfil}
                   #1
                   \par\nobreak
                  }

\def\subsection#1\par{\bigbreak
                      \global\advance\subsecno by 1  
                      \noindent
                      \hbox to \secindent{\the\secno.\the\subsecno\hfil}
                      #1
                      \par\nobreak
                     }

\def\appendix#1\par{\bigbreak
                    \noindent
                    APPENDIX #1
                    \par\nobreak
                   }

\newcount\equationno
\equationno=0
 
\def\enumber#1{\global\advance\equationno by 1
               $$#1
               \eqno(\the\equationno)
               $$}
 
\def\ealignnumber#1{\global\advance\equationno by 1
                    $$\openup.5\jot
		    \eqalignno{#1
                    &(\the\equationno)}
                    $$}

\newcount\refno
\refno=0

\def\cite#1{\global\advance\refno by 1%
 {\footnote{$^{\the\refno}$\nobreak}
           {\baselineskip=\footbaselineskip%
            \let\rm=\footrm \let\bf=\footbf \let\sl=\footsl \let\it=\footit%
            \frenchspacing\footrm #1\nonfrenchspacing}}%
 } 

\newdimen\citespace
\citespace=1pt

\def\citecomma{\kern-\citespace\nobreak{}$^,$}

\newcount\figno
\figno=0

\def\figure#1#2{\global\advance\figno by 1
                \topinsert
                #1
		\baselineskip=\figurebaselineskip
                \noindent
                FIGURE \the\figno. \ #2
		\vglue 12pt
                \endinsert}

\def\boxparameters{
\hsize=\boxhsize \parindent=0.5cm \parskip=0.2cm 
\pretolerance=1000 \tolerance=5000
\baselineskip=\boxbaselineskip
\abovedisplayskip=6pt \belowdisplayskip=6pt 
\abovedisplayshortskip=3pt \belowdisplayshortskip=4pt
                  }

\def\contrast#1#2#3{
\setbox\leftbox=\vtop{\boxparameters
\centerline{\bf Classical physics}
#2}
\setbox\rightbox=\vtop{\boxparameters
\centerline{\bf Quantum physics}
#3}
\setbox\wholebox=\hbox to \hsize{\hglue\quarterextra
                                 \box\leftbox
                                 \hglue\quarterextra
                                 \vrule width\rulewidth
                                 \hglue\quarterextra
                                 \box\rightbox
                                 \hglue\quarterextra
                                } 
\vskip 18pt plus 8pt minus 6pt
\bigbreak
\vbox{\centerline{\bf #1}
      \medskip
      \box\wholebox
     }
\vskip 18pt plus 8pt minus 8pt
                       }

\def\finishcontrast#1#2{
\setbox\leftbox=\vtop{\boxparameters #1}
\setbox\rightbox=\vtop{\boxparameters #2}
\setbox\wholebox=\hbox to \hsize{\hglue\quarterextra
                                 \box\leftbox
                                 \hglue\quarterextra
                                 \vrule width\rulewidth
                                 \hglue\quarterextra
                                 \box\rightbox
                                 \hglue\quarterextra
                                } 
\vbox{\box\wholebox}
\vskip 18pt plus 8pt minus 8pt
                      }
\def\probvec#1{\tilde #1}

\def\underrel#1#2{\mathrel{\mathop{#1}_{#2}}}


\def\subcap#1{_{\mkern-2.5mu\lower.35ex\hbox{$\scriptstyle#1$}}}
\def\subsubcap#1{_{\mkern-2mu\lower.25ex\hbox{$\scriptscriptstyle#1$}}}
\def\sm{\mkern 2mu}
\def\lg{\log_2\kern-0.05em}
\def\artbit{225\sm000}
\def\artdec{67\sm500}
\def\spinhalf{spin-$1\over2$} 
\def\zero{{\tt 0}}
\def\one{{\tt 1}}
\def\zeroket{|{\tt 0}\rangle}
\def\oneket{|{\tt 1}\rangle}
\def\tenbits{\one\one\zero\one\zero\zero\zero\one\one\zero}

\def\sA{{\cal A}}

\def\sJ{{\cal J}}
\def\sJc{{\cal J}_{\rm cl}}
\def\sJq{{\cal J}_{\rm q\kern-0.5pt u}}
\def\sN{{\cal N}}
\def\sS{{\cal S}}
\def\sV{{\cal V}}
\def\vc{\Delta v_{\rm cl}}
\def\vq{\Delta v_{\rm qu}}
\def\pspoint{Q_1,\ldots,Q_F,P_1,\ldots, P_F}
\def\trace{{\rm tr}\mkern 1mu}

\def\ppsi{p\bigl(|\psi\rangle\bigr)}

\def\ket{\rangle}
\def\bra{\langle}

\newbox\leftbox
\newbox\rightbox
\newbox\wholebox

\headline={\ifnum \pageno=1 \hfil \else\hss\tenrm\folio\hss\fi}
 
\line{}
\vskip 1.5cm
\centerline{QUANTUM INFORMATION:}
\centerline{HOW MUCH INFORMATION IN A STATE VECTOR?\footnote{$^*$}
{\baselineskip=\footbaselineskip\footrm 
Supported in part by the U.S.\ Office of Naval Research (Grant 
No.~N00014-93-1-0116).}}

\vskip 1.5cm

\begingroup
\baselineskip=\figurebaselineskip

\centerline{Carlton M.~Caves and Christopher A.~Fuchs\footnote{$^\dagger$}
{\baselineskip=\footbaselineskip\footrm 
Thanks to H.~Barnum, R.~Jozsa, R.~Schack, and B.~Schumacher for 
enlightening discussions.  CAF thanks G.~Comer in particular for
extensive correspondence concerning the point of view set forth here.}}
\centerline{Center for Advanced Studies, Department of Physics and Astronomy,}
\centerline{University of New Mexico, Albuquerque, New Mexico 
\thinspace 87131-1156, USA.}

\vskip 1.25cm
\parindent=1.3cm

\centerline{ABSTRACT}
{\narrower\smallskip\noindent
Quantum information refers to the distinctive information-processing 
properties of quantum systems, which arise when information is stored
in or retrieved from nonorthogonal quantum states.  More information 
is required to prepare an ensemble of nonorthogonal quantum states 
than can be recovered from the ensemble by measurements.  Nonorthogonal 
quantum states cannot be distinguished reliably, cannot be copied or 
cloned, and do not lead to exact predictions for the results of 
measurements.  These properties contrast sharply with those of 
information stored in the microstates of a classical system.  
\bigskip}
\endgroup

\section INTRODUCTION
 
The last fifteen years have seen a steadily increasing exchange of 
ideas between physicists and information theorists.  Physicists have 
become interested in how modern ideas of information processing affect 
the physical description of the world around us, and computer scientists
and communication theorists have become interested in fundamental 
questions of how physical law affects information processing.  The 
most fruitful new ideas have arisen from applying information theory 
to quantum physics, because information in quantum physics is radically 
different from classical information.

To begin the discussion, we need only the most primitive notion of 
information content.  The fundamental unit of information content, 
the {\it bit}, involves two alternatives, conveniently labeled \zero\ 
and \one.  A bit is not a physical system: it is the abstract unit of 
information that corresponds to two alternatives; its use implies
nothing about how the information is embodied in a physical medium.  
To transmit ten bits of information from us to you, we assemble a 
ten-bit message---a string of ten \zero s and \one s, say 
\tenbits---and send it to you.  Information thus has to do with 
selecting one possibility out of a set of alternatives; moreover,
information content is a logarithmic measure of the number of 
alternatives.  In the ten-bit example, we select a particular string
from the $\sN=2^{10}=1\sm024$ possible strings, so the information 
content is $\lg\sN=10$ bits.  This article, containing nominally about 
$\artbit$ bits,%
\cite{This estimate is obtained by multiplying the number of characters 
in the \TeX\ file for this article by 2.14 bits/character; see 
discussion in Sec.~2.}  
corresponding to $2^{\artbit}\simeq10^{\artdec}$ possible articles, 
is a more ambitious attempt to transmit information.

One can appreciate the difference between classical and quantum 
information by comparing and contrasting the physical realizations 
of a bit in classical and quantum physics.  The classical realization 
of a bit is a classical system that has two possible states---for 
example, a piece of paper that can have either a \zero\ or a \one\ 
written on it.  To send a bit from us to you, we send you the piece
of paper, and you examine it.  We provide one bit of information to 
{\it specify\/} which one-bit message to write on the paper or, 
putting it differently, to {\it prepare\/} the appropriate piece 
of paper.  You {\it acquire\/} this one bit of information when 
you examine the paper and determine which message we sent.  The 
paper, while in transit, can be said to {\it carry\/} the one bit 
from us to you. The defining feature of classical information is 
that when we send an $N$-bit message, by preparing one of $2^N$ 
alternatives, you can acquire all $N$ bits, by distinguishing among 
the alternatives.  This feature is {\it not\/} an automatic
consequence of physical law.  Rather, it is a consequence of using 
a classical medium to carry the information: in classical physics 
you are able to distinguish any alternatives we can prepare. 

The quantum realization of a bit is a two-state quantum system---for 
example, a \spinhalf\ particle.  A \spinhalf\ particle can be used 
to send one bit of classical information---and no more than one 
bit---encoded in two orthogonal states, e.g., spin ``down'' for 
\zero\ and spin ``up'' for \one.  Thus it is convenient to denote 
the state of spin ``down'' by $\zeroket$ and the state of spin ``up'' 
by $\oneket$.  The difference between classical and quantum two-state
systems is that quantum-mechanical superposition gives a quantum 
two-state system other possible states, not available to a classical 
two-state system: any linear combination of $\zeroket$ and $\oneket$ 
is also a possible state.  For a \spinhalf\ particle these states are 
in one-to-one correspondence with directions of the particle's spin.  

The crucial distinction between quantum and classical information
appears when one attempts to use these other states as alternatives 
for transmitting information.  Suppose we attempt to encode ten bits 
of information onto a \spinhalf\ particle, by preparing it so that 
its spin points in one of $2^{10}$ possible directions.  Then we send 
the particle to you.  Can you read out the ten bits?  Of course not. 
Quantum theory forbids any measurement to distinguish all $1\sm024$ 
possibilities.  Indeed, our over-enthusiasm in trying to stuff ten 
bits into the particle means that the amount of information you can 
recover is considerably less than a bit.  Nevertheless, there is 
a sense in which the \spinhalf\ particle actually does carry ten 
bits from us to you; if we want to transmit a description of 
the particle's state, so that you can prepare another \spinhalf\ 
particle with spin in the same direction, we must send you ten 
classical bits of information.  The ten bits of information needed 
to specify the particle's state, stored in some way in the particle, 
but not accessible to observation, are an example of quantum information.  
The fundamental difference between the information-storage and 
information-retrieval properties of classical and quantum two-state 
systems has been recognized by dubbing a quantum two-state system 
a {\it qubit}.%
\cite{B. Schumacher, ``Quantum coding,'' {\sl Phys. Rev.~A\/} {\bf 51}(4),
2738--2747 (1995).}\citecomma%
\newcount\refSchumacherPRA \refSchumacherPRA=\refno
\cite{{\it Qubit\/} is a shorthand for the minimal quantum system, 
a two-state quantum system, that can carry a bit of information.  
Logically, if one wishes to give a special name to the minimal 
physical system that can carry a bit, one should do so for both 
classical and quantum two-state systems, calling them perhaps c-bits 
and q-bits.  We are reluctant to use the neologism ``qubit,'' because
it has no standard English pronunciation, the ``qu'' being pronounced
as in ``cue ball,'' instead of as in ``queasy.'' We prefer ``q-bit,''
but acquiesce in the use of ``qubit,'' which has attained a degree of
general acceptance.}

It is worth repeating the qubit example in a general context.  A 
quantum system can encode classical information in a set of orthogonal 
states, because orthogonal states can be reliably distinguished by 
measurements.  The number of orthogonal states is limited by the 
dimension $D$ of the system's Hilbert space, and hence the maximum 
amount of classical information the system can carry is $\lg D$ bits.  
To emphasize what this means, suppose the system consists of $N$ 
qubits, where we use $N=4$ as a running example.  The dimension of 
Hilbert space, $D=2^N=16$, is exponentially large in the system size 
$N$, and the maximum classical information content, $\lg D=N=4$ bits, 
is given by the system size.

The superposition principle implies that any linear combination of 
orthogonal states is a possible state, so the number of possible states 
of the quantum system is arbitrarily large, limited only by how 
precisely one specifies the complex quantum-mechanical probability 
amplitudes for the chosen orthogonal states.  Suppose that one gives 
each amplitude to $m$-bit accuracy ($m/2$ bits each for the real and 
imaginary parts), where we use $m=10$ bits as a running example.  Only 
$D-1$ amplitudes need be specified, because one amplitude, first made 
real by a choice of overall phase, is then fixed by normalization.  
Thus the information needed to specify a state---the quantum information 
content of the state---is $m(D-1)$ bits, far larger than the $\lg D$ 
bits of classical information; likewise, the number of quantum states, 
$2^{m(D-1)}$, is far larger than the Hilbert-space dimension $D$. For 
the example of $N=4$ qubits, the quantum information, 
$m(D-1)=m(2^N-1)=150$ bits, is exponentially large in system size, and 
the number of states, $2^{m(D-1)}=2^{m(2^N-1)}\simeq10^{45}$, is larger 
by yet another exponential.  {\it Hilbert space is gratuitously big}---much 
bigger than the space needed to carry $\lg D$ bits.  Yet, because no 
measurement can distinguish all these states, almost none of this huge 
amount of information is accessible to observation.

How much information is in a state vector?  A heck of a lot, but
almost none---seemingly a paradox.  Fortunately, we stand to gain much 
from this circumstance: in the words of John Wheeler, ``No progress 
without a paradox!''%
\cite{J.~A.~Wheeler, ``From Mendel\'{e}ev's atom to the collapsing
star,'' in {\it Philosophical Foundations of Science}, edited by
R.~J.~Seeger and R.~S.~Cohen (Reidel, Dordrecht, 1974), pp.~275--301.}
The paradoxical character of quantum information is the impetus behind 
work in the fledgling field of {\it quantum information theory\/}, a field
seeking to elucidate the nature of quantum information, to quantify it
in meaningful ways, and to discover ``senses'' in which the enormous
information content of quantum states can be used.

This article is an introduction to the nature of quantum information.
Section~2 begins with the notion of information as having to with 
selecting one alternative from an ensemble of possibilities and shows 
how to quantify the information content of an ensemble in terms of the 
Gibbs-Shannon information measure.  Section~3 sets the stage for the 
remainder of the article by precisely defining the fundamental 
alternatives, called ``microstates,'' for physical systems, in both 
classical and quantum physics.  Classical microstates are fine-grained
cells on phase space, and quantum microstates are normalized state
vectors, or pure states, in Hilbert space.  Section~4 contrasts two 
measures of the information content of an ensemble of microstates: 
``preparation information'' is the information required to prepare 
a physical system in a particular microstate drawn from the ensemble, 
and ``missing information'' is the information that must be acquired 
from a measurement to place the system in a microstate.  Preparation 
information and missing information are identical for classical systems, 
but can be quite different for quantum systems.  Sections~5 through 7 
focus on three closely related information-theoretic 
concepts---predictability, distinguishability, and clonability---that 
strike at the heart of the distinction between ensembles of classical 
and quantum microstates.  Classical microstates lead to precise 
predictability for all measurements, can be distinguished with certainty, 
and can be copied or cloned precisely.  Nonorthogonal quantum pure states, 
in contrast, have none of these properties.  Section~8 closes the article 
by noting that these clean information-theoretic distinctions disappear 
when one compares ensembles of quantum pure states not with ensembles 
of classical microstates, but with ensembles whose alternatives are 
overlapping probability distributions for classical microstates.  
Section~8 thus provides motivation for a longer article, currently in 
preparation, which explores subtle aspects of quantum information that 
arise in comparing quantum pure states with classical probability
distributions.  This article serves as a starting point for the longer
paper.

Much of this article is devoted to making a distinction between 
``maximal'' and ``complete'' information about physical systems.  
In classical physics maximal information is complete.  The distinctive
feature of quantum physics is that maximal information is never 
complete, there being no way to obtain complete information about 
a quantum system.  The distinction between maximal and complete 
information in quantum physics was brought to the fore in the 
historic paper of Einstein, Podolsky, and Rosen, a paper that even 
now, after 60 years, inspires and challenges our thinking.  We humbly 
dedicate this contribution to the memory of Nathan Rosen.

\section INFORMATION CONTENT: A PRIMER

Before proceeding to a comparison of classical and quantum information, 
we need to sharpen up the primitive notion of information content 
used in the Introduction, where the information content of $\sN$ 
alternatives was $\lg\sN$ bits.  No probabilities appear here, yet 
it is clear that they ought to.  Suppose, for example, that only two 
of the alternatives can really occur, those two alternatives being 
equally likely.  There being effectively only two alternatives, the 
information content of the $\sN$ alternatives is only one bit, not 
$\lg\sN$ bits.  

To get at the notion of information content, we thus must first 
consider probabilities.  Throughout this article we adopt the Bayesian 
view of probabilities,%
\cite{B.~de Finetti, {\sl Theory of Probability}, 2 volumes (Wiley,
Chichester, 1974/75).}\citecomma%
\cite{L.~J. Savage, {\sl The Foundations of Statistics}, 2nd Ed.  (Dover, 
New York, 1972).}\citecomma%
\cite{E.~T. Jaynes, {\sl Papers on Probability, Statistics and Statistical
Physics}, edited by R.~D. Rosenkrantz (Kluwer, Dordrecht, 1983).}\citecomma%
\cite{J.~M. Bernardo and A.~F.~M. Smith, {\it Bayesian Theory\/} 
(Wiley, New York, 1994).}\citecomma%
\newcount\refBayestext \refBayestext=\refno%
\cite{E.~T. Jaynes, {\it Probability Theory: The Logic of Science},
to be published.}\ 
\newcount\refJaynestext \refJaynestext=\refno
which holds that a probability is a measure of credible belief based 
on one's state of knowledge.  The Bayesian view is particularly compelling 
in an information-theoretic context.  Probabilities are assigned to a set 
of alternatives based on what one knows, on one's stock of information 
about the alternatives.  These probabilities are often called 
{\it ignorance probabilities}: ``One of the alternatives actually occurs, 
but since I don't know which, all I can do is assign probabilities based 
on what I do know.''  Throughout this article we call the existing 
stock of information, used to make a probability assignment, 
{\it prior\/} information.  In defining information content, we seek 
a measure of {\it additional\/} information beyond the prior information 
that went into assigning the probabilities.  What is this additional 
information?  It is the further information, given the prior information, 
required to {\it prepare\/} a particular alternative, or it is the 
further information, given the prior information, that is 
{\it acquired\/} when one determines a particular alternative.  

The task of translating a state of knowledge into a probability 
assignment is the subject of Bayesian probability theory.  Among the
chief accomplishments of this theory are a set of rules for assigning 
prior probabilities in certain cases where the prior information can 
be given a precise mathematical formulation and the standard rule, 
called Bayes's rule, for updating a probability assignment as one 
acquires new information.  The general problem of translating a state 
of knowledge into a probability assignment is, however, far from 
solved and is the focus of an exciting field of contemporary research.%
$^{\the\refBayestext,\the\refJaynestext}$\citecomma%
\cite{See also the collection of MaxEnt conference proceedings under
the common title {\it Maximum Entropy and Bayesian Methods\/} (Kluwer,
Dordrecht).}\
That it is not completely solved does not concern us in this article.

Suppose then that there are $\sJ$ alternatives, labeled by an index
$j$, and that alternative $j$ has probability $p_j$.  We call a
collection of alternatives together with their probabilities an
{\it ensemble}.  To exhibit the role of probabilities in information 
content, consider what we call a {\it Gibbs ensemble}, an imaginary 
construct in which the ensemble of alternatives is repeated $N$ times,
$N$ becoming arbitrarily large.  The possible configurations of the
Gibbs ensemble are sequences of $N$ alternatives, there being $\sJ^N$ 
sequences in all.  As $N$ becomes large, the probability for the 
occurrence frequencies of the various alternatives becomes concentrated 
on those freqencies that match the probabilities in the original ensemble.
Thus we only need to consider those sequences for which the frequency 
of each alternative matches its probability.  The total number of such 
sequences is given by the multinomial coefficient
\enumber{
\sN=
{N!\over\displaystyle{\prod_{j=1}^\sJ(Np_j)!}}
\;,}
\newcount\eqGibbsno \eqGibbsno=\equationno
and each occurs with probability
\enumber{
\sN^{-1}=
\prod_{j=1}^\sJ p_j^{Np_j}
\;,}
\newcount\eqGibbspr \eqGibbspr=\equationno
where Stirling's formula relates (\the\eqGibbsno) to (\the\eqGibbspr) 
for large $N$.  Hence the information content of the Gibbs 
ensemble---the information required to prepare a particular sequence 
or the information acquired when one determines a particular 
sequence---is $\lg\sN=NH$, where
\enumber{
H\equiv
-\sum_{j=1}^\sJ p_j\lg p_j
}
\newcount\eqGSinfo \eqGSinfo=\equationno
is called the {\it Gibbs-Shannon information}.%
\cite{C.~E. Shannon, ``A mathematical theory of communication,'' 
{\sl Bell Syst. Tech. J.} {\bf 27}, 379--423 (1948) (Part~I); 
{\bf 27}, 623--656 (1948) (Part~II).  Reprinted in book form, 
with postscript by W.~Weaver: C.~E. Shannon and W.~Weaver, 
{\sl The Mathematical Theory of Communication\/} (University of 
Illinois, Urbana, IL, 1949).}\
Where it is helpful to indicate explicitly the dependence of the
Gibbs-Shannon information on a particular probability distribution, 
we denote it by $H(\probvec p)$, the symbol $\probvec p$ standing for the 
entire distribution.

The Gibbs-Shannon information $H(\probvec p)$ can be {\it interpreted\/} 
as the {\it average\/} information content per member of the Gibbs 
ensemble, i.e., as the average information content of the original 
ensemble.  It is the average information required to {\it specify\/} a
particular alternative within the original ensemble or the average 
information {\it acquired\/} when one determines a particular alternative
within the ensemble.  For $\sJ$ alternatives the Gibbs-Shannon 
information ranges from zero to $\lg\sJ$.  When the prior information 
determines a particular alternative, one assigns unit probability to 
that alternative and zero probability to all the rest, which leads to 
$H=0$; this is the sensible result that when one alternative is known 
definitely to occur, no information is acquired when it is determined.  
When the prior information does not discriminate at all among the 
alternatives, one assigns a uniform probability distribution, which 
leads to the maximum value of $H=\lg\sJ$; the discussion in the 
Introduction thus corresponds to assuming minimal prior information 
and, consequently, a uniform probability distribution.  Generally, 
the Gibbs-Shannon information measures the ignorance that leads to 
a probability assignment: $H(\probvec p)$ is the amount of information 
required to remove the ignorance expressed by the probabilities $p_j$.  

It is interesting to speculate how these considerations affect our
estimate that this article conveys $\artbit$ bits of information.  
That estimate assumes an average information per letter of 2.14 bits, 
considerably less than $\lg$(number of different English 
letters)$\hbox{}\simeq5$ bits.  The reason is that different English 
letters and combinations of letters occur with markedly different 
frequencies.  The figure of 2.14 bits/letter takes into account 
correlations between neighboring letters in English text by counting 
word frequencies for the $8\sm727$ most common English words, 
calculating an information per word, and dividing by an average word 
length (including spaces) of 5.5 letters to get an information per 
letter.%
\cite{C.~E. Shannon, ``Prediction and entropy of printed English,''
{\sl Bell Syst. Tech. J.} {\bf 30}, 50--64 (1951).}
To estimate the information content of the present text, we multiply
the number of characters in the \TeX\ file by 2.14; in doing so, we
are ignoring the difference between letters and \TeX-characters.

This article is correlated from beginning to end; the correlations 
cannot be ignored in assessing its information content.  Even with 
the nearby correlations of English taken into account, almost all of 
the $2^{\artbit}\simeq10^{\artdec}$ alternative articles corresponding 
to $\artbit$ bits are gibberish over scales longer than a few words. 
There is zero probability that we would compose them, and the 
reader---more importantly, the editors!---would assign zero probability 
for them to appear in this volume.  Of the remaining, much smaller 
amount of information, some conveys the essential ideas, but most has 
to do with our style and with attempts to make the essential ideas 
accessible.  In any case, we {\it must\/} leave to the reader the 
delicate task of estimating the essential information conveyed by 
this article, for that depends critically on the reader's prior knowledge.

An example of making information accessible, at the cost of redundancy, 
is a pause to summarize: in the Bayesian view a probability assignment 
incorporates what one knows about a set of alternatives; the 
Gibbs-Shannon information quantifies the additional information, 
beyond what one already knows, to pin down a particular alternative.

\section MICROSTATES AND STATES

The remainder of this article formulates differences between 
classical and quantum information.  We draw sharp distinctions 
between information in classical physics and information in quantum 
physics by translating into information-theoretic language things 
physicists already know.  We deliberately make the discussion 
detailed, the risk of tedium being outweighed, we hope, by two 
benefits.  First, many physicists are unfamiliar with information 
theory and, hence, have trouble appreciating information-theoretic 
concepts; how better to gain an appreciation of information theory 
than to see it in action on familiar ground?  Second, the sharp 
distinctions drawn in this article point to subtler questions that 
arise in distinguishing classical probability distributions from 
quantum state vectors; these questions, crucial to distinguishing 
quantum from classical information, are not considered in this 
article, but are taken up in a subsequent paper.

We are interested in the storage of information in and retrieval of 
information from physical media and, in particular, in the differences 
between classical and quantum media.  Since information has to do 
with picking one alternative out of a set of possible alternatives, 
we must spell out the fundamental alternatives, called {\it microstates}, 
in classical and quantum physics.    

The arena for classical physics is phase space.  At any time a 
classical system is located at a point in phase space; its dynamics 
traces out a path through phase space.  For specificity, we assume 
that the system of interest is described on a $2F$-dimensional phase
space, equipped with canonical co\"ordinates $\pspoint$; where it is
convenient, we abbreviate a phase-space point to $X=\bigl(\pspoint\bigr)$.
Furthermore, we assume that the accessible region of phase space has a 
finite volume 
\enumber{
\sV\subcap{F}=
\sA^F
\;,}  
where $\sA$ is a typical phase-space area per pair of canonical 
co\"ordinates.

The fundamental alternatives in classical physics are phase-space
points.  Yet we cannot specify a typical phase-space point, since 
the information required to do so is infinite.  To keep this information 
finite, we imagine that there is a finest scale on phase space.  This 
finest scale is characterized by a ``resolution volume'' 
$\vc=h_0^F\ll\sV\subcap{F}$, where $h_0$ is a resolution area per pair of 
canonical co\"ordinates.  We grid phase space into fine-grained cells 
of uniform volume equal to the resolution volume.  At this level of 
fine graining, the fundamental alternatives for a classical system---the 
classical microstates---are these fine-grained cells.  The microstates 
can be labeled by an index $j$, and the $j$th microstate can be 
specified by the phase-space address $X_j=\bigl(\pspoint\bigr)_j$ 
of, say, its central point.  The number of classical microstates at this 
level of fine graining is 
\enumber{
\sJc=
{\sV\subcap{F}\over\vc}=
\left({\sA\over h_0}\right)^{\!F}
\;.}

Turn now to quantum physics, where the dynamics unfolds within the
arena of Hilbert space.  More precisely, the relevant space is the 
space of Hilbert-space rays---normalized state vectors, with vectors 
that differ by a phase factor considered to be equivalent---a space 
called {\it projective Hilbert space}.  The dynamics of a quantum 
system traces out a path on projective Hilbert space.  We assume
throughout that Hilbert space is finite-dimensional, and we let $D$
denote the number of dimensions.

The fundamental alternatives in quantum physics are normalized state 
vectors, but the information required to specify a typical state vector 
is infinite.  To keep this information finite, we again imagine that 
there is a finest resolution, this time on projective Hilbert space.  
To define a notion of resolution on Hilbert space, we use the natural 
measure of distance on projective Hilbert space.  This natural 
distance between state vectors $|\psi\ket$ and $|\psi'\ket$ is 
the {\it Hilbert-space angle\/}%
\cite{W.~K. Wootters, ``Statistical distance and Hilbert space,''
{\sl Phys. Rev.~D\/} {\bf 23}, 357--362 (1981).}
\enumber{
\phi\equiv
\cos^{-1}\!\bigl(\big|\bra\psi|\psi'\ket\bigr|\bigr)
\;.}
Hilbert-space angle translates the overlap between quantum states into 
a distance function that is derived from a Riemannian metric on 
projective Hilbert space, called the Fubini-Study metric.%
\cite{J.~Anandan and Y.~Aharanov, ``Geometry of quantum evolution,'' 
{\sl Phys. Rev. Lett.} {\bf 65}(14), 1697--1700 (1990).}\citecomma%
\cite{J.~Anandan, ``A geometric approach to quantum mechanics,'' 
{\sl Found. Phys.} {\bf 21}(11), 1265--1284 (1991).}\citecomma%
\cite{G.~W. Gibbons, ``Typical states and density matrices,'' 
{\sl J. Geom. Phys.} {\bf 8}, 147--162 (1992).}\citecomma%
\newcount\refGibbons \refGibbons=\refno%
\cite{S.~L. Braunstein and C.~M. Caves, ``Statistical distance and the
geometry of quantum states,'' {\sl Phys. Rev. Lett.} {\bf 72},
3439--3443 (1994).}\  
The angle $\phi$ ranges from zero, when 
$|\psi'\ket=e^{i\alpha}|\psi\ket$, to a maximum value of $\pi/2$,
when $|\psi\ket$ and $|\psi'\ket$ are orthogonal.

The volume element $d\sV\subcap{D}$ induced by the Fubini-Study metric
is put in a form convenient for our purposes by Schack, D'Ariano,
and Caves.%
\cite{R.~Schack, G.~M. D'Ariano, and C.~M. Caves, ``Hypersensitivity to
perturbation in the quantum kicked top,'' {\sl Phys. Rev.~E\/} 
{\bf 50}(2), 972--987 (1994).}\ 
\newcount\refSchackDArianoCaves \refSchackDArianoCaves=\refno
They choose a fiducial state vector $|\psi_0\ket$ (analogous to the 
north pole in three real dimensions) and write an arbitrary normalized 
state vector as 
\enumber{
|\psi\ket=\cos\phi\,|\psi_0\ket+\sin\phi\,|\eta\ket
\;.}
Here $\phi\le\pi/2$ is a ``polar'' angle, the Hilbert-space angle 
between $|\psi\ket$ and $|\psi_0\ket$, the phase freedom in 
$|\psi\ket$ has been removed by choosing 
$\bra\psi|\psi_0\ket=\cos\phi$ to be real and nonnegative, 
and $|\eta\ket$ is a normalized vector in the ($D-1$)-dimensional 
subspace orthogonal to $|\psi_0\ket$.  An integral over projective 
Hilbert space, i.e., an integral over $|\psi\ket$, can be 
accomplished by integrating over the the polar angle $\phi$ and 
over the ($2D-3$)-dimensional sphere of normalized vectors 
$|\eta\ket$.  The volume element takes the 
form$^{\the\refSchackDArianoCaves}$
\enumber{
d\sV\subcap{D}=
(\sin\phi)^{2D-3}\cos\phi\,d\phi\,d\sS\subcap{\mkern 1mu 2D-3}
\;,}
\newcount\eqqVD \eqqVD=\equationno
where $d\sS\subcap{\mkern 1mu 2D-3}$ is the standard ``area'' element on a 
($2D-3$)-dimensional unit sphere.  This form of the volume element
is most useful for integrands that are symmetric about the fiducial
vector and thus depend only on the polar angle $\phi$.  Integrating 
over all of projective Hilbert space yields a total 
volume$^{\the\refGibbons,\the\refSchackDArianoCaves}$
\enumber{
\sV\subcap{D}=
\int d\sV\subcap{D}=
\sS\subcap{\mkern 1mu 2D-3}\int_0^{\pi/2}d\phi\,(\sin\phi)^{2D-3}\cos\phi=
{\sS\subcap{\mkern 1mu 2D-3}\over2(D-1)}=
{\pi^{D-1}\over(D-1)!}
\;.}
Here $\sS\subcap{\mkern 1mu 2D-3}=2\pi^{D-1}/(D-2)!$ is the area of a 
($2D-3$)-dimensional unit sphere.

To characterize the finest level of resolution on projective Hilbert 
space, we introduce a quantum ``resolution volume'' $\vq\ll\sV\subcap{D}$.  
It is convenient to think of these resolution volumes as tiny spheres 
whose radius, in terms of Hilbert-space angle, is $\phi\ll1$.  The 
volume of a sphere of radius $\phi$ is$^{\the\refSchackDArianoCaves}$ 
\enumber{
\vq=
\sS\subcap{\mkern 1mu 2D-3}\int_0^\phi d\phi'\,(\sin\phi')^{2D-3}\cos\phi'=
(\sin\phi)^{2(D-1)}\sV\subcap{D}\simeq\phi^{2(D-1)}\sV\subcap{D}
\;,} 
where the last form holds for the tiny spheres contemplated here.
We assume that the resolution volumes are small enough that sums 
over resolution volumes can be freely converted to integrals over 
projective Hilbert space.  

At this level of resolution on Hilbert space, the fundamental
alternatives---the quantum microstates---are the resolution 
volumes.  The quantum microstates can be labeled by an index 
$j$, and the $j$th microstate can be represented by the state vector 
$|\psi_j\ket$ that lies at the center of the $j$th sphere.%
\cite{A sphere is properly represented not by a state vector, but 
by a uniform distribution of state vectors within the sphere; for small
resolution volumes, the difference is unimportant.  The analogue in 
classical physics is that a fine-grained cell is properly represented 
not by its central point, but by a uniform probability density on the 
cell.}\   
A microstate can be specified, for example, by the probability 
amplitudes $\bra n|\psi_j\ket$ of the state vector in a 
specific orthonormal basis $|n\ket$, $n=1,\ldots,D$, i.e., by the 
expansion $|\psi_j\ket=\sum_{n=1}^D |n\ket\bra n|\psi_j\ket$.  
The number of quantum microstates at this level of resolution is%
\cite{C.~M. Caves, ``Information, entropy, and chaos,'' in {\sl
Physical Origins of Time Asymmetry}, edited by J.~J. Halliwell, 
J.~P\'erez-Mercader, and W.~H. Zurek (Cambridge University, Cambridge, 
England, 1994), pp.~47--89.}
\newcount\refCavesSpain \refCavesSpain=\refno
\enumber{
\sJq=
{\sV\subcap{D}\over\vq}=
\phi^{-2(D-1)}
\;.}

That there is in practice a finest level of resolution in the description 
of a physical system follows ineluctably from the inability to store or 
to process infinite amounts of information.  The size of this finest 
level of resolution, in classical or quantum physics, might be set 
by indifference to distinctions on yet finer scales or by physical 
constraints---e.g., by the resolution of laboratory equipment that is 
available for manipulating the system of interest.  It doesn't really 
matter how the finest scale is chosen; our discussion relies only on
choosing a finest scale, not on its actual size. 

A microstate is a state at the finest level of description.  To say 
that the system occupies a particular microstate requires that one 
have {\it maximal\/} information about the system.  Microstates are 
thus the states that are determined by maximal information.  Though
the term ``microstate'' does convey the notion of a state at the
finest level of description, it fails to convey what for us is the 
more important idea, that of a state specified by maximal information.  
Nevertheless, lacking a better term, we stick with ``microstate'' to 
conform to standard physics terminology.

\contrast{Microstates: states specified by maximal information}
{
A microstate is a fine-grained cell
\enumber{
X_j=
\bigl(\pspoint\bigr)_j
}
in phase space.
}
{
A microstate is a normalized state vector
\enumber{
|\psi_j\ket=
\sum_{n=1}^D |n\ket\bra n|\psi_j\ket
}
in Hilbert space.
}

What if one does not have maximal information about the system?  Then, 
according to the Bayesian view, one assigns probabilities $p_j$ to the 
microstates based on what one {\it does\/} know.  These probabilities
quantify ignorance about which microstate the system occupies.  The 
resulting ensemble of microstates and probabilities we call a 
{\it state\/} of the system.  The word ``state'' thus denotes an 
ensemble in the particular case that the ensemble's alternatives are
microstates of a physical system.  We stress that {\it the system
state depends on 
what one knows about the system}.  A microstate is a special case of 
a state---the special case that is specified by maximal information, 
i.e., by knowledge of the fundamental alternative.  In quantum physics 
a microstate (or state vector) is often called a {\it pure state}, 
whereas an ensemble in which more than one state vector has nonzero 
probability is called a {\it mixed state}.

For a classical system the ensemble of microstates and 
probabilities---the classical state---is equivalent to a phase-space 
probability density 
\enumber{
\rho(X)=
\sum_{j=1}^{\sJc}p_j\rho_j(X)
\;,}
\newcount\eqpsd \eqpsd=\equationno
where $\rho_j(X)$ is the normalized uniform density on the $j$th 
fine-grained cell.  For a quantum system the ensemble of microstates and 
probabilities---the quantum state---gives rise to a {\it density 
operator\/}
\enumber{
\hat\rho=
\sum_{j=1}^{\sJq}p_j|\psi_j\ket\bra\psi_j|=
\int d\sV\subcap{D}\,\ppsi|\psi\ket\bra\psi|
\;.}
\newcount\eqdo \eqdo=\equationno
The last form comes from converting the sum to an integral over 
projective Hilbert space, with $\ppsi$ being a probability density 
on projective Hilbert space.  

A quantum density operator is sufficient to predict the statistics of 
all measurements made on the system.  Consider, for example, what we
call a pure von Neumann measurement, the only kind of measurement 
considered in this article.  A pure von Neumann measurement is a
measurement of a nondegenerate observable---i.e., a nondegenerate 
Hermitian operator---and can be described completely by an orthonormal 
measurement basis $|n\ket$, $n=1,\ldots,D$, the eigenbasis of the
measured Hermitian operator.  (More generally, von Neumann measurements 
are described in terms of {\it orthogonal\/} projection operators, 
which project onto orthogonal Hilbert-space subspaces; the measurements 
here are called {\it pure\/} because they are described by 
{\it one-dimensional\/} orthogonal projectors $|n\ket\bra n|$, the 
projectors onto the orthogonal pure states $|n\ket$.)  Given the 
ensemble of state vectors $|\psi_j\ket$ and probabilities $p_j$, the 
probability for a pure von Neumann measurement to yield result $n$ is
\enumber{
q_n=
\sum_{j=1}^{\sJq}|\bra n|\psi_j\ket|^2 p_j=
\bra n|\hat\rho|n\ket=
\trace\bigl(\hat\rho\,|n\ket\bra n|\bigr)=
\sum_{m=1}^D|\bra n|\phi_m\ket|^2\lambda_m
\;.}
\newcount\eqqn \eqqn=\equationno
The first form here is a conventional probability formula, since
$|\bra n|\psi_j\ket|^2$ is the conditional probability to obtain 
result $n$, given that the system has state vector $|\psi_j\ket$.  
The second form in (\the\eqqn) introduces the density operator 
$\hat\rho$ of (\the\eqdo) and shows that it contains the statistics 
of all pure von Neumann measurements.  The third form, with the 
probability written in terms of a trace, is a form that can be 
extended to more general kinds of measurements.  The last form 
follows from expanding $\hat\rho$ in terms of its own orthonormal 
eigenbasis $|\phi_m\ket$, 
\enumber{
\hat\rho=
\sum_{m=1}^D \lambda_m|\phi_m\ket\bra\phi_m|
\;,}
\newcount\eqdodecomp \eqdodecomp=\equationno
where $\lambda_m$ is the eigenvalue of $\hat\rho$ associated with
the eigenvector $|\phi_m\ket$.  The expansion of a density operator 
in terms of its own eigenstates and eigenvalues is called its 
{\it orthogonal\/} (or spectral) {\it decomposition.}  The eigenvalues 
$\lambda_m$ make up a normalized probability distribution; indeed, if 
the measurement basis is chosen to be the eigenbasis $|\phi_m\ket$, 
$\lambda_m$ is the probability to obtain result $m$.

Though the density operator $\hat\rho$ is sufficient to predict the 
statistics of all measurements, it is unlike a classical phase-space
density in that it is {\it not\/} equivalent to the system state, 
i.e., to the ensemble of microstates and probabilities.  Many 
different ensembles give rise to the same density operator.  
Hughston, Jozsa, and Wootters%
\cite{L.~P. Hughston, R.~Jozsa, and W.~K. Wootters, ``A complete
classification of quantum ensembles having a given density matrix,''
{\sl Phys. Lett.~A\/} {\bf 183}, 14--18 (1993).}
\newcount\refHughston \refHughston=\refno
have outlined a procedure for constructing all ensembles that 
lead to a given density operator.  The lack of equivalence between
states and density operators is particularly important when a system 
can be divided into subsystems.  Suppose, for example, that a system 
is made up of two subsystems, and suppose that, having maximal 
information about the composite system, we assign it a state vector 
$|\Psi\ket$.  This joint state vector can be expanded as 
\enumber{
|\Psi\ket=\sum_m\sqrt{\lambda_m}|\phi_m\ket|\eta_m\ket
\;,}
where the state vectors $|\phi_m\ket$ are orthogonal state vectors
of subsystem~1 and the state vectors $|\eta_m\ket$ are orthogonal
state vectors of subsystem~2.  This kind of expansion of a joint pure 
state is called the {\it Schmidt decomposition}.%
\cite{A.~Peres, {\sl Quantum Theory: Concepts and Methods\/} 
(Kluwer, Dordrecht, 1993).}\ 
\newcount\refPeresbook \refPeresbook=\refno

The statistics of all measurements on one of the subsystems, say 
subsystem~1, can be derived from the {\it marginal density operator}
for that subsystem,
\enumber{
\hat\rho_1\equiv
\trace_2\bigl(|\Psi\ket\bra\Psi|\bigr)
\;,}
where $\trace_2$ denotes a {\it partial trace\/} over subsystem~2.  
Any operator $\hat O$ on the joint system can be expanded in terms
of a product basis $|n\ket|k\ket$,
\enumber{
\hat O=\sum_{n,k,n',k'}O_{nk,n'k'}|n\ket|k\ket\bra n'|\bra k'|
\;,} 
where the vectors $|n\ket$ are an orthonormal basis in the Hilbert
space of system~1 and the vectors $|k\ket$ are an orthonormal basis 
in the Hilbert space of system~2.  A partial trace over subsystem~2
yields an operator on subsystem~1, defined by
\enumber{
\trace_2(\hat O)\equiv
\sum_k\bra k|\hat O|k\ket=
\sum_{n,n'}\biggl(\sum_kO_{nk,n'k}\biggr)|n\ket\bra n'|
\;.}
Carrying out the partial trace to find $\hat\rho_1$, we arrive at
\enumber{
\hat\rho_1=
\sum_{m,m'}\sqrt{\lambda_m\lambda_{m'}}\,
|\phi_m\ket\bra\phi_{m'}|\,
\trace_2\bigl(|\eta_m\ket\bra\eta_{m'}|\bigr)
=\sum_m\lambda_m|\phi_m\ket\bra\phi_m|
\;.}
The states $|\phi_m\ket$ and the coefficients $\lambda_m$ are thus 
the eigenstates and eigenvalues of $\hat\rho_1$.  The marginal density 
operator $\hat\rho_1$ is a useful tool for deriving the statistics of 
measurements on subsystem 1, but {\it there is no justification for 
regarding $\hat\rho_1$ as associated with any specific ensemble of 
state vectors and probabilities for subsystem~1}.  In particular, one 
should not regard the state of subsystem~1 as being the ensemble of 
eigenvectors of $\hat\rho_1$ with eigenvalue probabilities.  The state 
in this situation is defined at the level of the joint system; there 
is no state, in our language, for subsystem~1 alone.

At times throughout the rest of this article we use the example of the 
{\it uniform ensemble}, for which the probabilities $p_j$ are all the 
same.  For this case, the classical density $\rho(X)=1/\sV\subcap{F}$ 
is uniform on the accessible region of phase space.  Similarly, the 
probability density $\ppsi=1/\sV\subcap{D}$ is uniform on projective 
Hilbert space, and by the symmetry of this ensemble, the quantum 
density operator~(\the\eqdo) is a multiple of the unit operator
$\hat 1$ on the $D$-dimensional Hilbert space:  
\enumber{
\hat\rho=
\int{d\sV\subcap{D}\over\sV\subcap{D}}\,|\psi\ket\bra\psi|=
{\hat 1\over D}
\;.} 
This ensemble has the virtue of highlighting most dramatically the 
distinctions between classical and quantum information. 
  
\section PREPARATION INFORMATION AND MISSING INFORMATION 

For the remainder of this article, we contrast classical and 
quantum information by investigating the storage of information 
in and retrieval of information from classical and quantum systems.  
The conceit we adopt is the one used in the Introduction: {\it we\/} 
prepare a system in a particular microstate drawn from an ensemble of 
microstates labeled by~$j$, which have probabilities $p_j$, and then 
{\it we\/} send the system to {\it you\/} for your examination.  
{\it We\/} know which microstate the system occupies; {\it we\/} 
must provide information to prepare the system in that microstate.  
{\it You}, not knowing which state we prepared, ascribe to the 
system the state described by microstate probabilities $p_j$.

The average amount of information we must provide to {\it prepare\/} 
the system in a particular microstate is the Gibbs-Shannon information
$H(\probvec p)$ corresponding to the probabilities $p_j$. This is
also the amount of information required to {\it specify\/} a particular 
microstate within the ensemble of microstates.  We call such information 
{\it preparation information}, or {\it specification information}, and 
denote it by $I$.  We stress that this preparation information is 
{\it not\/} the information needed to prepare the ensemble, i.e., the 
system state that has microstate probabilities $p_j$; rather, given 
the system state, it is the average information needed to prepare 
or to specify a particular microstate within the ensemble.  Classically 
the preparation information can be written as an integral over phase 
space, and quantum mechanically it can be written as an integral 
over projective Hilbert space:  
\enumber{
I=
H(\probvec p)=
-\sum_j p_j\lg p_j=
\cases{
\displaystyle{-\int d\sV\subcap{F}\,\rho(X)\lg\bigl(\rho(X)\vc\bigr)}\;,
&classical,\cr
\noalign{\vglue 6pt}
\displaystyle{-\int d\sV\subcap{D}\,\ppsi\lg\Bigl(\ppsi\vq\Bigr)}\;,
&quantum.\cr}
}
\newcount\eqprepinfo \eqprepinfo=\equationno

How might we prepare a particular microstate?  One way, which works
both classically and quantum mechanically, is to start with the 
system in a standard microstate and then to apply a specially 
designed Hamiltonian that causes the system to evolve into the desired 
microstate over a specified time interval.  For this purpose one can 
imagine a complicated apparatus that manipulates the system.  This
preparation apparatus has a dial, whose settings correspond to the 
system microstates.  Setting the dial to the $j$th microstate adjusts 
the system Hamiltonian to the designer Hamiltonian that causes the 
system to evolve into the $j$th microstate.  The setting for the $j$th 
microstate is to be used with probability $p_j$.  The information we 
must provide to pick a particular dial setting is the preparation 
information $I$. \ Parkins {\it et al.}%
\cite{A.~S. Parkins, P.~Marte, P.~Zoller, and H.~J. Kimble, 
``Synthesis of arbitrary quantum states via adiabatic transfer
of Zeeman coherence,'' {\sl Phys. Rev. Lett.} {\bf 71}(19), 3095--3098 
(1993).}
have proposed an example of this sort of procedure for preparing state
vectors of an electromagnetic field mode (a harmonic oscillator).

The designer-Hamiltonian method of state preparation highlights 
a crucial feature of the preparation information.  To prepare the
system in a particular microstate, we use another system---the 
preparation apparatus.  The preparation apparatus stores a 
record---its dial setting---of the prepared microstate.  {\it We}, 
knowing the dial setting, have maximal information and thus assign
the system a microstate; {\it you}, not knowing the dial setting, 
but knowing the microstate produced by each dial setting and the 
probabilities of the settings, ascribe to the system the state 
described by the possible microstates and their probabilities.  

The preparation information should be contrasted with the amount of 
information that you must acquire from a measurement to obtain maximal 
information about the system, i.e., to determine a system microstate.  
This amount of information that you are missing toward a maximal 
description of the system---{\it missing information\/} for short---is, 
as we see shortly, the {\it entropy\/}~$S$ of the state, measured in 
bits.

For a classical system there is no difference between preparation 
information and missing information.  You can make a measurement with 
the resolution of the fine-grained cells and thereby determine which
fine-grained cell the system occupies.  The average information you
acquire in such a measurement is $H(\probvec p)$, since microstate~$j$ 
occurs with probability $p_j$. The missing information is the same
as the information we provided to prepare the system---the preparation
information---and both are the same as the classical {\it entropy\/}
of the ensemble [cf.~(\the\eqprepinfo)].  

The contrast emerges in quantum physics.  If the ensemble of 
microstates includes, with nonzero probability, state vectors that 
are nonorthogonal, no measurement you make can determine which state 
vector we prepared.  In spite of this, a measurement {\it can\/} provide 
you maximal information about the system, but the state vector you 
ascribe to the system after your measurement---the one for which
the measurement provided maximal information---might not be included
in the original ensemble.  Here we are thinking about a pure von 
Neumann measurement, described by an orthonormal measurement basis 
$|n\ket$, $n=1,\ldots,D$; a pure von Neumann measurement provides 
maximal information by {\it leaving\/} the system in the basis state 
$|n\ket$ corresponding to the result $n$ of the measurement.

Given the ensemble of state vectors $|\psi_j\ket$ and probabilities 
$p_j$, the probability $q_n$ for your pure von Neumann measurement to 
yield result $n$ is given by~(\the\eqqn).  The average information 
you acquire from such a measurement is the Gibbs-Shannon information 
$H(\probvec q)$ corresponding to the probabilities $q_n$.  The
information you acquire clearly depends on the measurement basis 
$|n\ket$.  How then are we to identify a unique measure of missing 
information in quantum physics?  To do so, we return to our original 
question: how much information {\it must\/} you acquire to obtain 
maximal information about the system?  Thus we seek the von Neumann 
measurement that yields the {\it minimum\/} amount of information, 
for you must acquire at least this mimimal information to obtain a 
maximal description of the system.

To determine this quantum measure of missing information, we invoke a 
special property of the quantum conditional probabilities
$|\bra n|\phi_m\ket|^2$ that appear in (\the\eqqn).  The conditional 
probability $|\bra n|\phi_m\ket|^2$ has a dual character: for a 
measurement in basis $|n\ket$, it is the probability to obtain result 
$n$, given that the system has state vector $|\phi_m\ket$, and for
a measurement in basis $|\phi_m\ket$, it is the probability to obtain 
result $m$, given that the system has state vector $|n\ket$.  The 
special property$^{\the\refPeresbook}$ we need---dubbed, mysteriously, 
{\it double stochasticity}---is a straightforward consequence of this 
dual character and is simply that these conditional probabilities are 
normalized on both indices, $m$ and~$n$,
\enumber{
\sum_{n=1}^D|\bra n|\phi_m\ket|^2=
1=
\sum_{m=1}^D|\bra n|\phi_m\ket|^2
\;.}
Given this, we can write 
\ealignnumber{
H(\probvec q)-H(\probvec\lambda)&=
-\sum_{m,n}|\bra n|\phi_m\ket|^2\lambda_m\lg{q_n\over\lambda_m}\cr
&={1\over\ln2}\sum_{m,n}|\bra n|\phi_m\ket|^2
\!\left(-\lambda_m\ln{q_n\over\lambda_m}+(q_n-\lambda_m)\right)\ge
0
\;,}
\newcount\eqHdiff \eqHdiff=\equationno
where double stochasticity is used to insert the sum over 
$q_n-\lambda_m$, and where the inequality follows from the property 
$-\ln x\ge1-x$, for which equality holds if and only if $x=1$.  
Equality holds in (\the\eqHdiff) if and only if every term in the
second sum vanishes, i.e., $q_n=\lambda_m$ or $\bra n|\phi_m\ket=0$
for all $n$ and all $m$.  This necessary and sufficient condition for
equality is equivalent to $(\lambda_m-q_n)\bra\phi_m|n\ket=0$ for
all $n$ and $m$, which in turn is equivalent to
\enumber{
0=
\sum_m(\lambda_m-q_n)|\phi_m\ket\bra\phi_m|n\ket=
(\hat\rho-q_n)|n\ket
\hbox{\quad for all $n$.}
}
Thus equality holds in~(\the\eqHdiff) if and only if the measurement
basis $|n\ket$ is an eigenbasis of $\hat\rho$.

What we have shown is that$^{\the\refPeresbook}$\citecomma%
\cite{A.~Wehrl, ``General properties of entropy,'' {\sl Rev. Mod. Phys.}
{\bf 50}(2), 221--259 (1978).}
\newcount\refWehrl \refWehrl=\refno
\enumber{
H(\probvec q)\ge 
H(\probvec\lambda)=
-\sum_{m=1}^D\lambda_m\lg\lambda_m=
-\trace(\hat\rho\lg\hat\rho)\equiv
S(\hat\rho)
\;,}
where $S(\hat\rho)$, the {\it von Neumann entropy\/} of the state
(or ensemble), is determined by the density operator $\hat\rho$.  The 
von Neumann entropy plays a special role: it is the missing 
information---the minimum amount of information missing toward 
specification of a microstate---for any ensemble that has density 
operator $\hat\rho$.  The measurement that yields this minimum amount 
of information is a measurement in an eigenbasis of $\hat\rho$.  The 
von Neumann entropy ranges from zero, for a pure state, to a maximum 
of $\lg D$, for the density operator $\hat\rho=\hat 1/D$.  

In both classical and quantum physics, missing information, or entropy, 
$S$ is the amount of information missing toward a maximal description
of the system.  If $S=0$, there is no missing information; one already
knows which microstate the system occupies.  When $S$ is greater than
zero, there is a measurement that, by acquiring average information 
$S$, leaves the system in a microstate; no measurement that provides
less information than $S$ can leave one with maximal information about 
the system.%
\cite{These ideas are certainly not new with us.  Pauli, for instance, 
had similar thoughts on entropy: ``The first application of the calculus 
of probabilities in physics, which is fundamental for our understanding 
of the laws of nature, is the general statistical theory of heat, 
established by Boltzmann and Gibbs.  This theory, as is well known, 
led necessarily to the interpretation of the entropy of a system as 
a function of its state, which, unlike the energy, depends on our 
{\it knowledge} about the system.  If this knowledge is the maximal 
knowledge which is consistent with the laws of nature in general 
(micro-state), the entropy is always null.''  Quotation from W. Pauli, 
``Probability and physics,'' {\sl Dialectica} {\bf 8}, 112--124 (1954); 
translated in W.~Pauli, {\sl Writings on Physics and Philosophy}, 
edited by C.~P. Enz and K.~von Meyenn (Springer, Berlin, 1994).} 

We are now ready to appreciate the difference between preparation
information and entropy in quantum physics.$^{\the\refCavesSpain}$\citecomma%
\cite{C.~M. Caves, ``Information and entropy,'' {\sl Phys. Rev.~E\/}
{\bf 47}(6), 4010--4017 (1993).}
\newcount\refCavesPRE \refCavesPRE=\refno
It is obvious that the preparation information can be much bigger 
than the entropy.  The maximum value of the entropy is determined 
by the dimension $D$ of Hilbert space, i.e., by the number of 
{\it orthogonal\/} vectors that Hilbert space can accommodate, 
whereas the maximum value of the preparation information is 
determined by the number of state vectors, $\sJq$, in projective 
Hilbert space.  The number of state vectors is much larger than 
the number of orthogonal vectors, because any superposition of 
orthogonal state vectors is another state vector, and is limited 
only by one's resolution on projective Hilbert space; there is 
no corresponding situation in classical physics, because there 
is no way to combine two or more fine-grained cells to produce 
yet another fine-grained cell.  

The formal statement of the discrepancy between preparation information 
and entropy is that the former is never smaller than the latter,%
$^{\the\refHughston,\the\refWehrl}$\citecomma%
\cite{L.~B. Levitin, ``On the quantum measure of information,'' in
{\sl Proceedings of the Fourth All-Union Conference on Information 
and Coding Theory}, Sec.~II (Tashkent, 1969) (translation available
from A.~Bezinger and S.~L. Braunstein).  This paper has been essentially
reprinted as a part of L.~B. Levitin, ``Physical information theory
Part II: Quantum systems,'' in {\sl Workshop on Physics and Computation: 
PhysComp~'92}, edited by D.~Matzke (IEEE Computer Society, Los Alamitos, 
CA, 1993), pp.~215--219.}\citecomma%
\newcount\refLevitin \refLevitin=\refno
\cite{C.~M. Caves and P.~D. Drummond, ``Quantum limits on bosonic
communication rates,'' {\sl Rev. Mod. Phys.} {\bf 66}(2), 
481--537 (1994).}
\newcount\refCavesDrummond \refCavesDrummond=\refno
\enumber{
I=
H(\probvec p)=
-\sum_{j=1}^{\sJq}p_j\lg p_j\ge
-\trace(\hat\rho\lg\hat\rho)=
S(\hat\rho)
\;.}
\newcount\eqHpSrho \eqHpSrho=\equationno
Proofs of (\the\eqHpSrho) can be found in Refs.~\the\refHughston, 
\the\refWehrl, and \the\refCavesDrummond.  Equality holds in 
(\the\eqHpSrho) if and only if all the state vectors that have 
nonzero probability are orthogonal---i.e., if and and only if the 
state that gives rise to $\hat\rho$ is an ensemble of eigenstates 
of $\hat\rho$ with the probability of each eigenstate given by its 
eigenvalue.  

\contrast{Preparation information vs.\ entropy}
{
The amount of information, $I$, required to prepare a particular 
microstate within an ensemble of microstates is the same as the 
entropy $S$ of the ensemble.  For the uniform ensemble the amount 
of preparation information (or entropy) is
\ealignnumber{
I&=
\lg\sJc\cr
&=\lg(\sV\subcap{F}/\vc)\cr
&=F\lg(\sA/h_0)=
S
\;.\quad}
We can interpret $\lg(\sA/h_0)$ as the number of bits of preparation 
information per pair of canonical co\"ordinates.  
}
{
The amount of information, $I$, required to prepare a particular 
microstate within an ensemble of microstates is larger than the 
von Neumann entropy $S$ of the ensemble, unless the ensemble consists 
of orthogonal state vectors.  For the uniform ensemble the amount of 
preparation information is
\ealignnumber{
I&=
\lg\sJq\cr
&=\lg(\sV\subcap{D}/\vq)\cr
&=(D-1)\lg\phi^{-2}\gg
\lg D=
S(\hat\rho)\;.\cr
&
}
We can interpret $\lg\phi^{-2}$ as the number of bits of preparation 
information per probability amplitude (cf.~discussion in the 
Introduction; for the example there of 10 bits per amplitude, 
$\phi=1.79^\circ$).  
}

It is instructive at this point to compare directly the number of 
microstates for a system described in classical physics with the number 
of microstates for the {\it same\/} system described in quantum physics.  
To do so, imagine fine graining classical phase space on the quantum 
scale by choosing the resolution area per pair of canonical 
co\"ordinates, $h_0$, to be the Planck constant $h$.  The resulting 
number of quantum-scale fine-grained cells is $\sJc=\sV\subcap{F}/h^F$.  
If such a system is sufficiently classical, i.e., $\sJc\gg1$, then when 
the system is quantized, these quantum-level phase-space cells 
correspond roughly to {\it orthogonal\/} state vectors that span 
Hilbert space.  The number of quantum-level phase-space cells thus 
gives the dimension of Hilbert space, $\sJc=\sV\subcap{F}/h^F=D$.  The number 
of quantum microstates, $\sJq$, is exponentially larger,
\enumber{
\sJc=
D
\ll
2^{(D-1)\lg\phi^{-2}}=
\sJq
\;,}
as a direct consequence of quantum superposition: superposition of 
quantum-level phase-space cells produces an exponentially large number
of state vectors that have no classical counterpart.  Notice that 
this conclusion is true even if the resolution on projective Hilbert 
space is so coarse that it corresponds to giving only one bit per 
amplitude, i.e., $\lg\phi^{-2}=1$. 

One can see how quantum statistical physics manages to reduce to 
classical statistical physics in the classical limit, despite the far 
larger number of quantum microstates.  Statistical physics is founded 
on entropy, or missing information, not on preparation information.  
For a sufficiently classical system, the quantum density operator 
$\hat\rho=\sum_jp_j|\psi_j\ket\bra\psi_j|$ is approximately 
diagonal in an {\it orthonormal\/} basis of state vectors 
$|\psi_j\ket$ that can be identified with quantum-level cells on 
classical phase space.  The corresponding classical phase-space density 
is $\rho(X)=\sum_jp_j\rho_j(X)$, where $\rho_j(X)$ is the normalized
uniform density on the $j$th quantum-level cell.  Thus the von Neumann 
entropy reduces to the classical entropy, provided that the resolution 
on phase space is fixed at the quantum scale.

The information $H(\probvec q)$ you acquire from a pure von Neumann 
measurement provides the information you need to specify the system's 
state vector {\it after\/} the measurement, but for nonorthogonal 
ensembles the information you acquire is not sufficient to infer the 
state vector {\it before\/} the measurement, i.e., the state vector 
that we prepared.  For nonorthogonal ensembles, part of the information 
you acquire comes from the intrinsic unpredictability of quantum physics 
and tells you nothing about which state vector we prepared.  Indeed, 
for nonorthogonal ensembles, this useless part of the information is
always large enough that the part that is useful in determining which 
state vector we prepared, called the {\it accessible\/} information, 
is not just less than the preparation information $H(\probvec p)$, but 
is less than the von Neumann entropy $S(\hat\rho)$. 

At this point it is instructive to note that measurements themselves
provide another method of state preparation: observe a system, and 
thereby prepare it in the microstate corresponding to the result of the 
measurement.  In contrast to the use of designer Hamiltonians, however,
the state prepared by this method cannot be predicted in advance.  In 
considering preparation by measurements, assume for simplicity that 
the state of the system to be observed is the uniform ensemble.  
Classically, one measures which microstate the system occupies, 
thereby gathering the $\lg\sJc=S$ bits of missing information (or 
entropy), which coincides with the preparation information.  Things 
are different in quantum physics.  For the uniform ensemble, the 
density operator is a multiple of the unit operator.  Any orthonormal 
basis is an eigenbasis of this density operator, and thus any pure 
von Neumann measurement gathers $\lg D=S(\hat\rho)$ bits of missing 
information.  The rest of the preparation information, $\lg(\sJq/D)$ 
bits, comes not from the random measurement outcome, but from the 
selection of the measurement basis.  In contrast to classical physics, 
only a small part of the preparation information comes from observing 
the quantum system; most comes from choosing {\it how\/} to observe 
the system.%
\cite{More elaborate methods for preparing state vectors by measurements
have been considered by K.~Vogel, V.~M. Akulin, and W.~P. Schleich,
``Quantum state engineering'' {\sl Phys. Rev. Lett.} {\bf 71}(12),
1816--1819 (1993), and by B.~M. Garraway, B.~Sherman, H.~Moya-Cessa,
P.~L. Knight, and G.~Kurizki, ``Generation and detection of nonclassical
field states by conditional measurements following two-photon resonant 
interactions,'' {\sl Phys. Rev.~A\/} {\bf 49}(1), 535--547 (1994).}\ 

So far we have seen that preparation information and missing 
information are the same in classical physics, but can be quite 
different in quantum physics.  The difference is connected to the 
fundamental lack of predictability and distinguishability in quantum
physics.  Our objective is to sharpen up this connection by addressing 
three closely related questions, concerned with {\it predictability}, 
{\it distinguishability}, and {\it clonability}.%
\cite{The question of cloning, or copying, state vectors was first
considered by W.~K. Wootters and W.~H. Zurek, ``A single quantum cannot
be cloned,''{\sl Nature} {\bf 299}, 802--803 (1982), and independently
by D.~Dieks, ``Communication by EPR devices,'' {\sl Phys. Lett. A} 
{\bf 92}(6), 271--272 (1982).}\ 
\newcount\refcloning \refcloning=\refno
These questions are posed in terms of our conceit: {\it we\/} prepare
the system in a microstate drawn from an ensemble of microstates and 
probabilities and send the system to {\it you}; not knowing which 
microstate we prepared, {\it you\/} attribute to the system the state 
corresponding to the ensemble.

\section PREDICTABILITY

The first question concerns {\it predictability}: when one has 
maximal information about a system, do all measurements have 
predictable results?  The question, expressed in our conceit, becomes 
the following: {\it we\/} prepare the system in a microstate from 
the ensemble of microstates and probabilities and send the system 
to you; can {\it we\/} predict uniquely the result of any measurement 
{\it you\/} perform?  In both classical and quantum physics, the 
answer is easy.  For a classical system, if one knows the system's 
microstate, i.e., knows which fine-grained cell the system occupies, 
then one can predict the results of all measurements made on scales 
coarser than the chosen fine graining.  Measurements yield no new 
information.  In contrast, the essence of quantum physics is that 
even if one knows the system's microstate, i.e., knows its state 
vector, unpredictability remains.  The outcomes of most measurements 
are unpredictable and thus yield new information.

A convenient way to quantify the amount of new information was 
introduced by Wootters.%
\cite{W.~K. Wootters, ``Random quantum states,'' {\sl Found.~Phys.} {\bf
20}(11), 1365--1378 (1990).}\
\newcount\refWoottersFP \refWoottersFP=\refno
Given a particular state vector, pick at random a pure von Neumann 
measurement, and calculate the average information obtained from the 
measurement, the average being taken over the random choice of measurement.  
It is equivalent to reverse the roles of the state vector and the 
measurement$^{\the\refWoottersFP}$: start with a particular pure 
von Neumann measurement, described by an orthonormal 
measurement basis $|n\ket$, pick a random state vector, and then 
calculate the average information obtained from the measurement, the 
average being taken over the uniform ensemble of state vectors.  

If the state vector is $|\psi\ket$, the probability to obtain 
result $n$ is $|\bra n|\psi\ket|^2$, and the information 
obtained from the measurement is
\enumber{
H=
-\sum_{n=1}^D
|\bra n|\psi\ket|^2
\lg|\bra n|\psi\ket|^2
\;.}
\newcount\eqinfonpsi \eqinfonpsi=\equationno
Averaging over the randomly chosen vector $|\psi\ket$ yields an 
average information
\enumber{
\bar H=
-\sum_{n=1}^D
\int{d\sV\subcap{D}\over\sV\subcap{D}}\,
|\bra n|\psi\ket|^2
\lg|\bra n|\psi\ket|^2
\;.}
Every term in the sum is the same, since the integral is independent of
the basis vector $|n\ket$.  Replacing $|n\ket$ with a fiducial 
vector $|\psi_0\ket$ and using the volume element of~(\the\eqqVD), we 
can write the average information as
\ealignnumber{
\bar H&=
-D\int{d\sV\subcap{D}\over\sV\subcap{D}}\,
|\bra\psi_0|\psi\ket|^2
\lg|\bra\psi_0|\psi\ket|^2\cr
&=-D{\sS\subcap{\mkern 1mu 2D-3}\over\sV\subcap{D}}
\int_0^{\pi/2}d\phi\,(\sin\phi)^{2D-3}\cos^3\!\phi\,\lg(\cos^2\!\phi)\cr
&=-{D(D-1)\over\ln2}
\int_0^1 dx\,(1-x)^{D-2}x\ln x
\;.}
\newcount\eqbarH \eqbarH=\equationno
The final integral can be done by writing $x^s\ln x=(d/ds)x^s$, i.e.,
\enumber{
\int_0^1 dx\,(1-x)^{D-2}x\ln x=
{d\over ds}\!\left.\left(
\int_0^1 dx\,(1-x)^{D-2}x^s
\right)\right|_{s=1}=
-{1\over D(D-1)}\sum_{k=2}^D{1\over k}
\;,}
thus yielding an average information%
\cite{R.~Jozsa, D.~Robb, and W.~K. Wootters, ``Lower bound for accessible
information in quantum mechanics,'' {\sl Phys. Rev. A\/} {\bf 49}, 
668--677 (1994).}\citecomma%
\newcount\refJozsaRobbWootters \refJozsaRobbWootters=\refno
\cite{K.~R.~W. Jones, ``Entropy of random quantum states,'' 
{\sl J. Phys.~A\/} {\bf 23}(23), L1247--L1251 (1990); ``Riemann-Liouville
fractional integration and reduced distributions on hyperspheres,''
{\sl J. Phys.~A\/} {\bf 24}, 1237--1244 (1991).}
\newcount\refJonesJPA \refJonesJPA=\refno
\enumber{
\bar H=
{1\over\ln2}\sum_{k=2}^D{1\over k}
\;.}
\newcount\eqHbar \eqHbar=\equationno

For a two-dimensional Hilbert space, the average information, 
$\bar H=1/2\ln2=0.721$ bits, should be compared with the maximum
of 1 bit that can be obtained from a pure von Neumann measurement.
Similarly, for a three-dimensional Hilbert space, the average
information, $\bar H=5/6\ln2=1.202$ bits, should be compared to
the maximum of $\lg3=1.585$ bits. For large $D$, the asymptotic value
of the average information is
\enumber{
\bar H
\underrel{\sim}{D\gg1}
\lg D-{1-\gamma\over\ln2}=
\lg D-0.609\sm95
\;,}
where $\gamma=0.577\sm22$ is Euler's constant; this is just $0.610$ bits
shy of the maximum of $\lg D$ bits that can be obtained from a pure
von Neumann measurement.  In other words, even when one possesses
maximal information about a quantum system, the result of a typical 
pure von Neumann measurement is nearly completely unpredictable; the 
measurement yields almost the maximum amount of information that can 
be obtained from a pure von Neumann measurement.

\contrast{Predictability?}
{
{\bf Yes.} If one has the preparation information---i.e., one knows
which fine-grained cell the system occupies---then one can predict the
results of all measurements on scales coarser than the fine-grained
cells.  The amount of information acquired from any such measurement 
is zero.
}
{{\bf No.} If one has the preparation information---i.e., one knows the
system's state vector---one generally cannot predict the result of a 
measurement.  One acquires further information from a typical measurement.  
For a measurement chosen at random, the average amount of information 
acquired is 
\enumber{
\bar H=
{1\over\ln2}\sum_{k=2}^D{1\over k}
\underrel{\sim}{D\gg1}
\lg D-0.609\sm95\;\hbox{bits.}
}
}

The unpredictability of quantum physics lays bare one of its great
mysteries: one can gather an arbitrarily large amount of information 
from a quantum system, by making repeated pure von Neumann 
measurements in incompatible bases.  Where does all this information
come from?  A good example%
\cite{E.~P. Wigner, ``On hidden variables and quantum mechanical
probabilities,''{\sl Am. J. Phys.} {\bf 38}(8), 1005--1009 (1970). In
a footnote, Wigner attributes discussion of this example to von Neumann,
who based his private belief in the inadequacy of hidden-variable 
theories upon it.}
is provided by a \spinhalf\ particle whose spin is measured alternately 
along the $z$ and $x$ axes.  Each measurement yields a bit of 
information; these bits are plucked out of the system as though 
one were drawing from an inexhaustible well of 
information.$^{\the\refCavesPRE}$

How is this gathering of fresh information from repeated measurements
different from the fresh information that is acquired when a classical 
system is examined on successively finer scales?  In classical physics, 
if one knows which fine-grained cell a system occupies, unpredictability 
is solely a consequence of making measurements on a scale finer than 
the original fine graining.  If one determines which cell the system 
occupies at the new, finer scale, predictability is restored at that 
scale.  Not so in quantum physics.  Information gathered by repeated 
measurements has nothing to do with determining the system's state 
vector on finer and finer scales on projective Hilbert space.  The new 
information does not enhance predictability at all.  With each new
measurement some quantities become more predictable, while others
become less predictable, as in the example of the \spinhalf\ particle.  

What does this tell us about the status of probabilities in quantum 
physics?  Consider, for example, the density operator $\hat\rho$ 
of~(\the\eqdo) and (\the\eqdodecomp).  If one makes a measurement in 
the eigenbasis of $\hat\rho$, the probability to obtain result $m$ is 
given by the eigenvalue
\enumber{
\lambda_m=
\bra\phi_m|\hat\rho|\phi_m\ket=
\sum_{j=1}^{\sJq}|\bra\phi_m|\psi_j\ket|^2 p_j
\;.}
\newcount\eqlambda \eqlambda=\equationno
There appear to be two quite different kinds of probabilities in 
this expression: the prior probabilities $p_j$ express ignorance 
about the system's microstate; the conditional probabilities 
$|\bra\phi_m|\psi_j\ket|^2$, which give the probability to 
obtain result $m$ given that the system has state vector 
$|\psi_j\ket$, express the intrinsic unpredictability of quantum 
physics.  

One's first inclination is to view the conditional probabilities 
not as ignorance probabilities, but as something else, say, 
``quantum probabilities.''%
\cite{J.~von Neumann, ``Quantum logics (strict- and probability-logics),''
in {\sl John von Neumann: Collected Works}, Vol. IV, edited by A.~H.
Taub (Macmillan, New York, 1962), pp.~195--197.}\citecomma%
\cite{P.~Benioff, ``Possible strengthening of the interpretative rules
of quantum mechanics,'' {\sl Phys. Rev.~D\/} {\bf 7}(12) 3603--3609
(1973).}\citecomma%
\cite{M.~Strauss ``Two concepts of probability in physics,'' in
{\sl Logic, Methodology and Philosophy of Science IV}, edited by
P.~Suppes, L.~Henkin, A.~Jojo, and Gr.~C.~Moisil (North-Holland,
Amsterdam, 1973), pp.~603--615.}\citecomma%
\cite{K.~R. Popper, {\sl Quantum Theory and the Schism in Physics\/}
(Hutchinson, London, 1982).}
These ``quantum probabilities'' are determined by the rules of quantum 
physics; i.e., they are squares of Hilbert-space inner products.  Like 
the probabilities $|\bra n|\psi\ket|^2$ in (\the\eqinfonpsi), 
they can be thought of as conditional probabilities for measurement 
results, conditioned on a particular state vector, i.e., on a maximal 
description of the system.  How can they be expressions of ignorance,
when they follow from maximal information and the very laws of physics? 

Initial enthusiasm for having two different species of probabilities 
is dampened by the realization that they get hopelessly entangled, 
leaving no way to maintain a clean distinction.  In~(\the\eqlambda)
the probabilities $\lambda_m$ of the measurement outcomes are a
combination of the purported species of probabilities.  The form of 
the combination depends on the ensemble, even though the $\lambda_m$ 
themselves, the eigenvalues of $\hat\rho$, are blind to the ensemble 
that defines the density operator.  For example, if the ensemble 
consists of the eigenstates of $\hat\rho$, with probabilities 
$\lambda_m$, then these probabilities are purely ignorance probabilities, 
whereas if $\hat\rho$ is constructed from a nonorthogonal ensemble, 
the same probabilities $\lambda_m$ contain contributions from 
``quantum probabilities.'' 

We suggest that the best approach is to adhere to the Bayesian view, 
which holds that {\it all\/} probabilities are expressions of 
ignorance about a set of alternatives.  An argument%
\cite{For a related argument, see R.~N. Giere, ``Objective single-case 
probabilities and the foundations of statistics,'' in {\sl Logic, 
Methodology and Philosophy of Science IV}, edited by P.~Suppes, 
L.~Henkin, A.~Jojo, and Gr.~C.~Moisil (North-Holland, Amsterdam, 1973),
pp.~467--483.} 
against this approach---against viewing ``quantum probabilities'' as 
ignorance probabilities---runs as follows.  If probabilities express 
ignorance, then by removing the ignorance, one should be left with 
{\it complete\/} information, which permits one to predict everything 
with certainty.  In classical physics this is just what happens, but
in quantum physics it is not.  In classical physics one acquires 
complete information and complete predictability by determining which 
microstate the system occupies.  In quantum physics there is no 
procedure for removing all ignorance.  One is not allowed to acquire 
{\it complete\/} information; the best one can do is to acquire 
{\it maximal\/} information, which leaves one uncertain about the 
results of most observations.  Thus, the argument runs, since not 
all ignorance can be removed, ``quantum probabilities'' cannot be 
ignorance probabilities.

We reject this argument, because requiring that all ignorance be
removable is just a prejudice, not valid in quantum physics.  
``Quantum probabilities'' are ignorance probabilities; they express
ignorance about the outcomes of potential measurements.  {\it What 
is different in quantum physics is not the status of probabilities, 
but rather the nature of the alternatives.}  In classical physics, 
probabilities are concerned with {\it actualities}: ``One of the 
alternatives actually occurs, but since I don't know which, I assign 
probabilities based on what I do know.''  The probabilities that 
describe intrinsic quantum unpredictability---the ``quantum 
probabilities''---express ignorance about {\it potentialities}%
\cite{W.~Heisenberg, ``The development of the interpretation of the
quantum theory,'' in {\sl Niels Bohr and the Development of Physics},
edited by W.~Pauli (McGraw-Hill, New York, 1955), pp.~12--29;
W.~Heisenberg, {\sl Physics and Philosophy: The Revolution in
Modern Science\/} (Harper, New York, 1958).}
that are actualized by measurement: ``I know one of these alternatives 
{\it would\/} occur if I enquired about that set of alternatives, 
but since I don't know which, I assign probabilities based on what I 
do know.''  What is it that one knows? The system's state vector.  
Given that knowledge, quantum physics provides the rule for assigning 
probabilities to the results of {\it all\/} possible questions that 
can be addressed to the system.  The sometimes perceived weakness of 
Bayesianism---that there is no general theory for translating a state 
of knowledge into a probability assignment---does not apply to the 
case of maximal information in quantum physics.  Indeed, viewed in
this light, the quantum rule for assigning probabilities is the most 
powerful rule yet of Bayesian probability theory. 

Formally, one says that in classical physics, maximal information is
complete, but in quantum physics, it is not.%
\cite{It is worth noting for historical interest that neither Einstein 
nor Pauli would have been a stranger to this language, though they 
were in opposing camps on the foundations of quantum physics; their 
difference lay in their beliefs about the finality of this situation.  
In a letter from Pauli to Born, dated 1954 April 15, Pauli says, 
``What is more, on the occasion of my farewell visit to him [Einstein] 
he told me what we quantum mechanicists would have to say to make 
our logic unassailable (but which does {\it not} coincide with what 
he himself believes): `Although the description of physical systems 
by quantum mechanics is incomplete, there would be no point in 
completing it, as the complete description would not agree with the 
laws of nature.'\thinspace''  From M.~Born, {\sl The Born-Einstein 
Letters\/} (Walker, New York, 1971), p.~226.  Alternatively, in a 
letter from Pauli to M.~Fierz, dated 1954 August 10, Pauli says, 
``The famous `incompleteness' of quantum mechanics (Einstein) is 
somehow-somewhere really there, but of course cannot be remedied
by going back to the physics of classical fields (this is just a 
`neurotic misunderstanding' on Einstein's part.)\dots'' From K.~V. 
Laurikainen, {\sl Beyond the Atom: The Philosophical Thought of 
Wolfgang Pauli\/} (Springer, Berlin, 1988), p.~145 (translation
by R.~Schack).}\   
What should we demand of a physical theory in which maximal information 
is not complete?  Maximal information is a state of knowledge; the 
Bayesian view says that one must assign probabilities based on the 
maximal information.  Classical physics is an example of the special 
case in which all the resulting probabilities predict unique measurement 
results; i.e., maximal information is complete.  In a theory where 
maximal information is not complete, the probabilities one assigns on 
the basis of maximal information are probabilities for answers to 
questions one might address to the system, but whose outcomes are 
not necessarily predictable (some outcomes must be unpredictable, 
else the maximal information becomes complete).  This implies that the 
possible outcomes cannot correspond to actualities, existing objectively 
prior to asking the question; otherwise, how could one be said to have 
maximal information?%
\cite{Here we argue that in a theory where maximal information is
not complete, the quantities the theory deals with cannot all be 
actualities, i.e., objective properties.  Bell inequalities do something 
more in the case of quantum physics: they {\it demonstrate\/} that 
quantum physics has no extension to a theory in which maximal 
information is complete, i.e., a theory in which the statistical 
predictions of quantum physics are an expression of ignorance about 
underlying actualities, usually called hidden variables.  More 
precisely, the Bell inequalities show that the statistical predictions 
of quantum physics cannot be obtained from any {\it local\/} theory 
in which maximal information is complete.}\ 
Furthermore, the theory must provide a rule for assigning probabilities 
to {\it all\/} such questions; otherwise, how could the theory itself
be complete?  Quantum physics is consistent with these demands.  A more 
ambitious program would investigate whether the quantum rule is the 
{\it unique\/} rule for assigning probabilities in situations where 
maximal information is not complete.  You won't be surprised to learn 
that we don't know how to make progress on this ambitious program.  

With this perspective, let us return to the eigenvalue probabilities 
$\lambda_m$ in (\the\eqlambda).  These probabilities express ignorance 
about the result of a measurement in the eigenbasis of the density 
operator $\hat\rho$.  Different states of knowledge, i.e., different 
ensembles, can lead to the same density operator and thus to the same 
eigenvalue probabilities.  It is not the status of the probabilities 
$\lambda_m$ that changes when going from one such ensemble to another; 
what changes is the nature of the alternatives to which the probabilities 
apply.  For example, if the ensemble consists of the eigenstates of 
$\hat\rho$, with probabilities $\lambda_m$, then the alternatives have 
the properties of actualities, whereas if $\hat\rho$ is constructed from 
a nonorthogonal ensemble, the same alternatives can only be potentialities.  

The information gathered from repeated measurements on quantum systems 
is indeed drawn from an inexhaustible well, but it is a well of 
potentialities, not actualities.  Asked where all this information
resides, we reply, with apologies to Gertrude Stein%
\cite{Stein damned Oakland, California, her childhood home, with the 
comment, ``There is no there there.''}:
``There is no where there.''

\section DISTINGUISHABILITY

Putting aside this philosophical discussion, we return to our series 
of questions, the second of which concerns {\it distinguishability}: 
can microstates be distinguished reliably by measurements?  In terms 
of our conceit, the question becomes the following: {\it we\/} prepare
the system in a microstate from the ensemble of microstates and 
probabilities and send the system to {\it you}; can {\it you}, not 
knowing which microstate we prepared, determine the microstate from 
the result of a measurement?%
\cite{It is useful to spell out very clearly what is happening here.  
{\it We\/} prepare a system in such a way that we possess maximal 
information about it; that maximal information specifies a particular 
microstate from the given ensemble.  {\it We\/} send the system to you.  
The question is the following: can {\it you}, by performing measurements 
on the system, acquire the maximal information that {\it we\/} used to
prepare the system?  This formulation underscores the fact that your 
objective is to find out what we did---i.e., which of the possible 
procedures we used to prepare the system.  The system is the medium
from which you try to extract the information you desire.}
Again the answer is easy.  In classical physics, yes, because a 
measurement can determine which fine-grained cell we prepared.  In 
quantum physics, no, for nonorthogonal ensembles, because no 
measurement can distinguish nonorthogonal state vectors reliably.

The problem of trying to determine which microstate we sent---an 
{\it inference\/} problem---is easy to state, but when the inference 
is not completely reliable, it is difficult to formulate a quantitative 
measure of just how reliable the inference is.%
\cite{C.~A. Fuchs, {\it Distinguishability and Accessible Information
in Quantum Theory}, Ph.D. thesis, University of New Mexico (1996).}\
For this reason it is convenient to replace the inference problem with 
a related question taken from communication theory: {\it we\/} provide 
information to prepare the system in a microstate drawn from the 
ensemble of microstates and probabilities; can the preparation 
information be transmitted from us to {\it you}?  For a given 
measurement, the amount of information you acquire about which 
microstate we sent is called the {\it mutual information}. 
The mutual information is the amount of information transmitted 
from us to you; it cannot exceed the preparation information.

We are interested in your {\it optimal\/} measurement, the one that 
provides the most information about which microstate we sent.  The 
corresponding maximum value of the mutual information, called the 
{\it accessible information\/}%
\cite{B.~Schumacher, ``Information from quantum measurements,'' in
{\sl Complexity, Entropy, and the Physics of Information}, edited by 
W.~H. Zurek (Addison-Wesley, Redwood City, CA, 1990), pp.~29--37.}\
\newcount\refSchumacherSFI \refSchumacherSFI=\refno
and denoted here by $J$, provides a 
quantitative measure of the distinguishability of the microstates in 
the ensemble.  For an ensemble of classical microstates, you can read 
out all the preparation information by observing which fine-grained 
cell the system occupies.  Similarly, for an ensemble of 
{\it orthogonal\/} state vectors, you can read out all of the 
preparation information by making a pure von Neumann measurement 
in a basis that includes the orthogonal states.  For an ensemble of 
{\it nonorthogonal\/} state vectors, however, not all of the 
preparation information can be transmitted from us to you---the
accessible information is necessarily less than the preparation 
information---because no measurement you make can distinguish all 
the states in the ensemble. 

The uniform ensemble illustrates these points without requiring 
the entire formalism of quantum communication theory.  We select 
a state vector from the uniform ensemble and send it to you.  You 
make a pure von Neumann measurement described by an orthonormal 
basis $|n\ket$.  It doesn't matter which basis you choose, because 
any two bases are related by a unitary transformation, which leaves 
the uniform ensemble unchanged.  How much of the preparation 
information do you obtain?  All the measurement outcomes are equally 
likely, because the density operator is $\hat\rho=\hat 1/D$, so you 
acquire $\lg D$ bits from the measurement.  Yet almost all of this 
information is due to the intrinsic unpredictability of quantum 
mechanics---i.e., it is information about quantum potentialities that 
are actualized by the measurement---and thus it is not information 
about which state vector we sent.  Indeed, if you knew which state 
vector we transmitted, you would still obtain on average the 
unpredictability information $\bar H$ of (\the\eqHbar).  This portion 
of the $\lg D$ bits you obtain is evidently not information about 
which state vector we sent, so it must be subtracted from the 
$\lg D$ bits acquired from the measurement, to yield an accessible 
information 
\enumber{
J=
\lg D-\bar H=
\lg D-{1\over\ln2}\sum_{k=2}^D{1\over k}
\;.}

For a two-dimensional Hilbert space the accessible information is
$J=1-1/2\ln2=0.279$ bits, for a three-dimensional Hilbert space
it is $J=\lg3-5/6\ln2=0.383$ bits, and for large $D$ the asymptotic
value is
\enumber{
J
\underrel{\sim}{D\gg1}
{1-\gamma\over\ln2}=
0.609\sm95\;{\rm bits.}
}
The upshot is that you get almost none of the preparation information. 
Indeed, the accessible information is smaller than the von Neumann
entropy $S(\hat\rho)=\lg D$---much smaller for large dimensions---and
the von Neumann entropy, in turn, is much smaller than the preparation 
information, $I=(D-1)\lg\phi^{-2}$, for the uniform ensemble. 

\contrast{Distinguishability?}
{
{\bf Yes.} Measurements can distinguish different microstates 
unambiguously. Hence, the preparation information can be transmitted
reliably.  In the case of the uniform ensemble, the accessible 
information is
\enumber{
J=
\lg\sJc=
S=
I
\;;}
the accessible information is equal to the classical entropy $S$ and
to the preparation information $I$.
} 
{
{\bf No.} No measurement can distinguish nonorthogonal microstates
unambiguously.  Hence, not all the preparation information for a 
nonorthogonal ensemble can be transmitted reliably.  In the case 
of the uniform ensemble, the accessible information is
\ealignnumber{
J&=
\lg D-\bar H\cr
&=\lg D-{1\over\ln2}\sum_{k=2}^D{1\over k}
\underrel{\sim}{D\gg1}
0.609\sm95\;\hbox{bits;}\cr
&}
the information transmitted is smaller than the von Neumann 
entropy $S(\hat\rho)=\lg D$ (for $D\gg1$, much smaller), which in 
turn is much smaller than the preparation information $I$.
}

Jones%
\cite{K.~R.~W. Jones, ``Principles of quantum inference,'' {\sl Ann.
Phys. (N.Y.)\/} {\bf 207}(1), 140--170 (1991).}\citecomma%
\newcount\refJonesAnnals \refJonesAnnals=\refno
\cite{K.~R.~W. Jones, ``Fundamental limits upon the measurement of state
vectors,'' {\sl Phys. Rev.~A\/} {\bf 50}(5), 3682--3699 (1994).}\
\newcount\refJonesPRE \refJonesPRE=\refno
has formulated and investigated the problem of using measurements to 
determine which state vector is drawn from the uniform ensemble and
has generalized to the case where one is allowed many copies of the
same state vector.  He replaces the inference problem with the 
corresponding communication problem, just as we do here, and uses 
mutual information to characterize the inference power of the 
measurements.  As part of his investigation, he has developed 
powerful mathematical tools%
$^{\the\refJonesJPA,\the\refJonesPRE}$ 
for doing Hilbert-space integrals like~(\the\eqbarH).

In the example of the uniform ensemble, unpredictability translates 
directly into lack of distinguishability; i.e., the average information 
$\bar H$, which quantifies unpredictability, is subtracted from 
$\lg D$ to give the accessible information.  Intrinsic quantum 
unpredictability can be thought of as drawing from the well of 
information about potentialities.  Any attempt to distinguish 
nonorthogonal states must draw from this well, the resulting 
information acting as noise that defeats the attempt.  Though one 
can extract as much information as one wants from a quantum system, 
by going repeatedly to the well of information about potentialities, 
one cannot acquire the information needed to distinguish nonorthogonal 
state vectors. 

\section CLONABILITY

The third question on our list concerns {\it clonability}: can 
microstates be copied reliably?  In terms of our conceit, this 
question becomes the following: {\it we\/} prepare the system in
a microstate from the ensemble of microstates and probabilities 
and send the system to {\it you}; can {\it you}, not knowing which 
microstate we prepared, devise a procedure that, without changing 
the microstate of the system we sent, prepares a second system in 
the same microstate.  

The answer is easy this time partly because we can argue that clonability
is equivalent to distinguishability.  Distinguishability implies 
clonability: if microstates are distinguishable, you can determine 
which microstate we sent and then, employing an appropriate 
designer Hamiltonian, prepare a second system in the same 
microstate.  Conversely, clonability implies distinguishability
(provided you can determine a microstate if you have an arbitrarily 
large number of copies of it): if you can prepare one copy of a 
microstate, you can prepare an arbitrarily large number of copies 
and thereby determine the state.  The assumption here---that you
can determine a microstate if you have an arbitrarily large number
of copies---is certainly true in classical physics---the copies, though
unnecessary, don't hurt---but it is also true in quantum physics.  
Given an arbitrarily large number of copies of a state vector, you 
can identify the state vector by determining the statistics of pure 
von Neumann measurements in a sufficient number of incompatible bases.%
\cite{J.~L. Park and W.~Band, ``A general theory of empirical state
determination in quantum physics: Part I,'' {\it Found. Phys.} 
{\bf 1}(3), 211--226 (1971).}\citecomma%
\cite{I.~D. Ivanovi\'{c}, ``Geometrical description of quantal state
determination,'' {\sl J. Phys. A\/} {\bf 14}(1), 3241--3245
(1981).}\citecomma%
\cite{W.~K. Wootters, ``Quantum mechanics without probability
amplitudes,'' {\sl Found. Phys.} {\bf 16}(4), 391--405 (1986).}\

For an ensemble of classical microstates, you can observe which 
fine-grained cell the system occupies and then prepare a second system
in the same fine-grained cell.  Similarly, for an ensemble of 
{\it orthogonal\/} state vectors, you can determine which state
vector we prepared by making a pure von Neumann measurement in a 
basis that includes the orthogonal states and then prepare a second
system in the same state.  For an ensemble of {\it nonorthogonal\/} 
state vectors, however, you cannot devise a procedure that copies 
all the state vectors in the ensemble.

The standard demonstration$^{\the\refcloning}$ that state vectors 
generally cannot be cloned runs as follows.  Suppose one wishes to make
$N$ copies of a state vector $|\psi\ket$.  One starts with the original
system having state vector $|\psi\ket$ and with the $N$ systems 
that are to receive the copies having some standard state vector.
A copying transformation takes this initial state vector to a final 
product state vector in which all $N+1$ systems have state vector 
$|\psi\ket$:
\enumber{
|\psi\ket
\Biggl|
\matrix{\hbox{standard}\cr\hbox{state vector of}\cr\hbox{$N$ systems}\cr}
\Biggr\ket
\underrel{\longrightarrow}
{{\hbox{\sevenrm copying}\atop\hbox{\sevenrm transformation}}}
|\psi\ket
\underbrace{|\psi\ket\ldots|\psi\ket}_{\hbox{$N$ times}}
\;.}
There is nothing wrong with this transformation for a single state 
vector $|\psi\ket$, but problems arise when one tries to copy all the 
state vectors $|\psi_j\ket$ in an ensemble.  Since the transformation 
must be unitary, it must preserve the inner product between any two 
initial state vectors to which it is to be applied.  Thus unitarity 
requires that
\enumber{
\bra\psi_j|\psi_k\ket=
\bra\psi_j|\psi_k\ket
\Biggl\bra
\matrix{\hbox{standard}\cr\hbox{state vector of}\cr\hbox{$N$ systems}\cr}
\Biggm|
\matrix{\hbox{standard}\cr\hbox{state vector of}\cr\hbox{$N$ systems}\cr}
\Biggr\ket=
\bra\psi_j|\psi_k\ket^{N+1}
\;,}
which is equivalent to $\bra\psi_j|\psi_k\ket=0$ or 1.  The unitarity 
requirement can be met if and only if $|\psi_j\ket$ and $|\psi_k\ket$
are identical or orthogonal.  An ensemble of orthogonal state vectors 
can be cloned, but an ensemble of (distinct) nonorthogonal state vectors 
cannot.

This demonstration that nonorthogonal state vectors cannot be
cloned, when combined with the preceding argument that distinguishability 
implies clonability, seems to show definitively that nonorthogonal 
state vectors cannot be distinguished.  The alert reader, however, 
will wonder what this demonstration, which relies on unitarity, has to 
do with the preceding argument, which was couched in terms of 
measurements; perhaps nonorthogonal states can be cloned if one allows 
measurements to be part of the process.  It's easy to demonstrate that 
this cannot be the case.$^{\the\refcloning}$  An apparatus designed
to make measurements and to prepare copies can be included in the overall
copying transformation, which becomes a grand unitary transformation for 
the $N+1$ systems and the apparatus.  If the apparatus starts in some 
standard pure state, the copying transformation must take the form
\enumber{
|\psi\ket
\Biggl|
\matrix{\hbox{standard}\cr\hbox{state vector of}\cr\hbox{$N$ systems}\cr}
\Biggr\ket
\Biggl|
\matrix{\hbox{standard}\cr\hbox{state vector of}\cr\hbox{apparatus}\cr}
\Biggr\ket
\underrel{\longrightarrow}
{{\hbox{\sevenrm copying}\atop\hbox{\sevenrm transformation}}}
|\psi\ket
\underbrace{|\psi\ket\ldots|\psi\ket}_{\hbox{$N$ times}}
\Biggl|
\matrix{\hbox{final}\cr\hbox{state vector of}\cr\hbox{apparatus}\cr}
\Biggr\ket_{\mkern-6mu|\psi\ket}
\;,}
where the final apparatus state vector can depend on $|\psi\ket$.
The crucial feature of the final state---that it is a product 
state vector of the $N+1$ systems and the apparatus---is necessary 
because one desires a unique state vector for the $N+1$ systems.  
Carrying through the requirements of unitarity, one finds that 
including the apparatus does not change the previous conclusion.
 
\contrast{Clonability?}
{
{\bf Yes.} Microstates can be copied.
} 
{
{\bf No.} Microstates in an ensemble of nonorthogonal state vectors 
cannot be copied.
}

Why consider clonability separately from distinguishability when the 
two are equivalent?  The easy answer is that it's always good to have 
equivalent formulations of a problem.  In this case one can say more: 
the question of clonability is somehow easier to formulate than the 
question of distinguishability.  As a result, consideration of 
clonability allows one to see directly the principle that prevents 
cloning---and, hence, prevents distinguishing---nonorthogonal state 
vectors.  That principle is the unitarity of quantum physics.  No 
matter how involved the demonstrations of the inability to distinguish
nonorthogonal quantum states become, the underlying principle is 
unitarity.

\section CONCLUSION

This article has been devoted to comparing and contrasting the 
information storage and retrieval properties of classical and 
quantum systems.  Information is stored by preparing a system in 
a particular microstate drawn from an ensemble of possible microstates.  
Information is retrieved by observing the system and trying to 
determine which microstate was prepared.  The properties of 
classical information---i.e., information encoded in the microstates 
of a classical system---are a consequence of the distinguishability 
of classical microstates.  Information put into a classical system 
can be recovered by observing the system.  Information stored in 
orthogonal microstates of a quantum system acts just like classical 
information, because orthogonal state vectors can be distinguished 
by measurements.  Quantum information has something to do with the 
information needed to prepare or to specify a particular state vector
from an ensemble of nonorthogonal state vectors.  Though quantum 
measurements can generate as much new information as desired, the 
information used to prepare one state vector from an ensemble of 
nonorthogonal state vectors is not accessible to observation, because 
nonorthogonal state vectors cannot be distinguished.  

Should we stop here, satisfied with this clean distinction between 
classical and quantum microstates?  No, because the distinction 
disappears when one compares, instead, classical probability 
distributions and quantum state vectors.  An ensemble whose members 
are themselves {\it overlapping\/} probability distributions has none 
of the properties of classical information discussed in this article.  
Indeed, such an ensemble displays the contrasting properties of an 
ensemble of nonorthogonal state vectors: a probability distribution 
does not provide predictability, overlapping distributions cannot be 
distinguished, and overlapping distributions cannot be cloned.  Yet 
an ensemble of probability distributions is a purely classical concept 
and can have nothing to do with quantum information.  It must be 
possible---indeed, it is essential if we wish to understand the 
differences between classical and quantum information---to find clean 
distinctions between ensembles of probability distributions and 
ensembles of state vectors.  This task, for which this article serves 
as a prelude, we must defer to a more extensive future paper.

\bye